\documentclass[twocolumn,showpacs,preprintnumbers,amsmath,amssymb]{revtex4}

\usepackage{hyperref}
\usepackage{graphicx}
\usepackage{dcolumn}
\usepackage{bm}

\begin{document}


\title{Optical Rogue Waves in Whispering-Gallery-Mode Resonators}

\author{Aur\'elien Coillet$^1$}
\author{John Dudley$^1$}
\author{Go\"ery Genty$^2$}
\author{Laurent Larger$^1$}
\author{Yanne K. Chembo$^{1}$}
\affiliation{$^1$FEMTO-ST Institute [CNRS UMR6174], Optics Department, \\
        16 Route de Gray, 25030 Besan\c con cedex, France. \\
        $^2$Tampere University of Technology, Institute of Physics, \\
          Optics Laboratory, FIN-33101 Tampere, Finland}
\date{\today}

\begin{abstract}
We report a theoretical study showing that rogue waves can emerge in whispering
gallery mode resonators as the result of the chaotic interplay between Kerr
nonlinearity and anomalous group-velocity dispersion. The nonlinear dynamics of the
propagation of light in a whispering gallery-mode resonator is investigated
using the Lugiato-Lefever equation, and we evidence a range of parameters where 
rare and extreme events associated with a non-gaussian statistics of the
field maxima are observed.
\end{abstract}

\maketitle

Rogue (or ``freak'') waves are rare events of extreme amplitudes, that were
first investigated in hydrodynamics where their fascinating nature and
disastrous consequences have drawn noticeable attention. Further investigations
have shown that rogue waves may arise in various nonlinear physical systems,
including Bose-Einstein condensates, super-fluid helium 
and plasma waves~(see for example ref.~\cite{Review_rogue_waves}, and references therein). 
In optics, the first experimental observation
of rogue waves was reported in 2007~\cite{Solli2007}, and since then, optical
rogue waves have been the subject of extensive research activity in the context
of super-continuum generation or laser systems(see~\cite{Nature-peregrine,Kovalsky,Bonatto,Pisarchik,Arecchi,Review_John,SotoCrespo2011,Zaviyalov2012,Lecaplain2012} and references therein).
The theoretical modelling of optical rogue waves in optical fibers is based on the generalized
nonlinear Schr\"odinger equation (GNLSE), which includes various physical
phenomena such as dispersion (second order and higher), dissipation, as well as
Kerr and Raman nonlinearities. Here we provide theoretical evidence of rogue
waves in whispering-gallery mode resonators, resulting from the collision of
soliton breathers in the process of hyperchaotic Kerr comb generation.

\begin{figure}
  \begin{center}
    \includegraphics{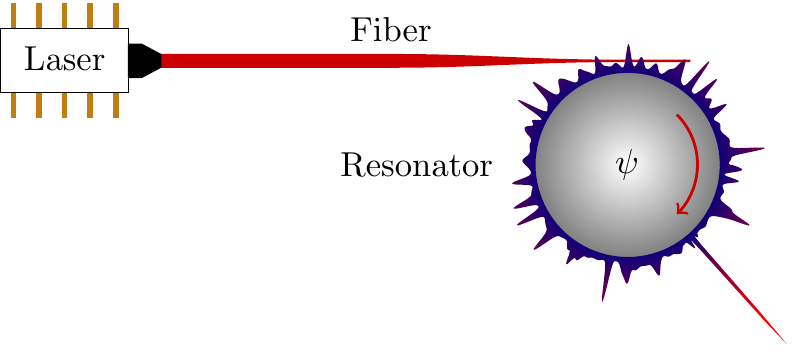}
  \end{center}
  \caption[Schema]{Schematic representation of the system under study. A
    continuous-wave laser radiation is coupled to a WGM cavity through the
    evanescent field of a tapered fiber. The WGM resonator can be a fused glass
    microsphere, a polished crystal disk, or an integrated resonator. Laser
    light is trapped by total internal reflection and travels along the
    azimuthally direction. The figure displays a snapshot of the numerical
    simulations that shows a rogue wave, which is in this case a sharp peak
    characterized by extreme amplitude and very rare occurrence.} 
  \label{fig:Schema}
\end{figure}

The system under study is schematically displayed in Fig.~\ref{fig:Schema}. A
continuous-wave laser at $1550$~nm is used to pump a whispering-gallery mode
resonator which traps the photons by total internal reflection. In these
resonators, the photon lifetime related to the resonance linewidth through
$\tau_{\rm ph} = 1/\Delta \omega$, and it can be significantly high (typically,
of the order of $1$~$\mu$s for a loaded quality factor of $Q \sim 10^9$).  In
the fundamental familly of eigenmodes (doughnut-like modes), the confined laser
radiation circulates in a torus inside the resonator. The eigenfrequencies of
the WGM resonator can be Taylor-expanded as $\omega_\ell = \omega_{\ell_0} +
\zeta_1 (\ell-\ell_0) +\frac{1}{2} \zeta_2 (\ell - \ell_0)^2$, where $\ell$ is
the eigennumber which unambiguously labels all the modes of the fundamental
family, $\omega_{\ell_0}$ is the eigenfrequency of the pumped mode $\ell =
\ell_0$, $\zeta_1 =\left. d\omega/d \ell\right|_{\ell=\ell_0} $ is the
free-spectral range of the resonator, and $\zeta_2 = \left. d^2\omega/d
\ell^2\right|_{\ell=\ell_0}$ is the second-order dispersion coefficient. 

It has been shown recently in refs.~\cite{Matsko,Chembo2013,Coen2013} that the spatiotemporal
dynamics of the intra-cavity laser field can be modeled using the
Lugiato-Lefever equation (LLE)~\cite{Lugiato1987}, which is a NLSE with damping,
driving and detuning: 
\begin{equation} 
  \frac{\partial\psi}{\partial \tau} = -(1+{i}\alpha)\psi + {i} |\psi|^2 \psi -
  {i} \frac{\beta}{2}\frac{\partial^2\psi}{\partial \theta^2} +F \, , 
  \label{eq:LLE}
\end{equation}
where $\psi$ is the normalized slowly-varying envelope of the intra-cavity
field, $\tau=t/2\tau_{\rm ph}$ is the dimensionless time, and $\theta \in
[-\pi,\pi]$ is the azimuthal angle along the circumference of the disk. The normalized
parameter $\alpha=-2(\Omega_0-\omega_{\ell_0})/\Delta \omega$ is scaling the physical detuning between the driving and the pumped eigenmode ($\Omega_0$ and
$\omega_{\ell_0}$, resp.), while the normalized parameter $\beta=-2\zeta_2/\Delta\omega$ is scaling the group-velocity dispersion (GVD) at the eigenmode frequency. We operate here in the regime of anomalous GVD which corresponds in the model to $\beta<0$. Finally, the square of
the dimensionless pump term $F$ in the LLE is proportional to the laser power
$P$ following:
\begin{equation}
  F^2 = \frac{8 g_0 }{ \hbar \Omega_0 } \frac{\Delta\omega_\text{ext}}{\Delta\omega^3} \, P  \, ,
  \label{eq:Power}
\end{equation}
where $g_0=n_2\text{c}\hbar\Omega_0^2/n_0^2V_0$ stands for the nonlinear gain,
$n_2$ is the nonlinear refractive index of the material, $V_0$ is the effective
volume of the pumped mode, and $\Delta\omega_\text{ext}$ is the contribution of the external coupling to the resonator linewidth. It is interesting to note that the intra-cavity field has been
normalized in a way that  $[\Delta \omega / 2g_0]\, |\psi|^2$ is the total
number of photons inside the resonator.
The LLE modeling in the photonic WGM context has already been compared successfully with Kerr comb generation
experiments~\cite{Coillet2013}, and is therefore a relevant theoretical tool to investigate the complex phenomenology of the laser field dynamics in the WGM resonator.

\begin{figure}
  \begin{center}
    \includegraphics{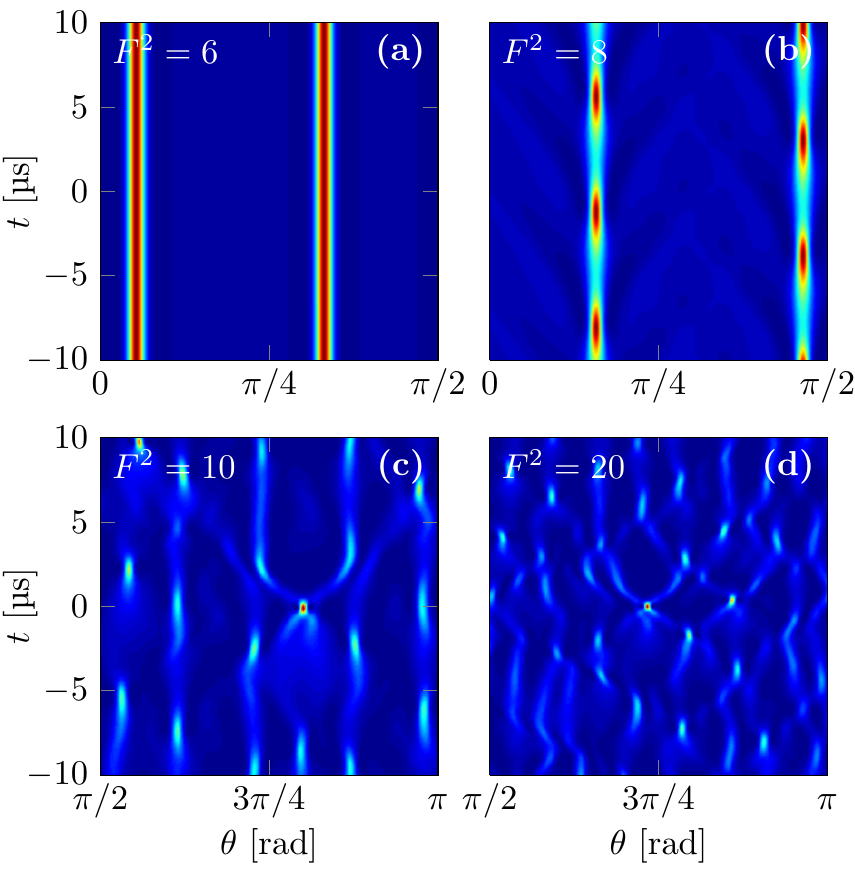}
  \end{center}
  \caption[Figure3]
  {Spatio-temporal representation of the evolution of the optical field
  in the WGM resonator. For $F^2=6$, stable solitons emerge from the
  noisy initial condition, and remain stable. For $F^2=8$, soliton breathers are
  obtained. As the pump is increased ($F^2=10$ and $20$), their number increases and they start to interact, thereby leading to unexpected, rare and extremely high amplitude waves. Note that only a fraction of the total
  cavity circumference is represented for better visibility.}
  \label{fig:pcolors}
\end{figure}
\begin{figure}
  \begin{center}
    \includegraphics{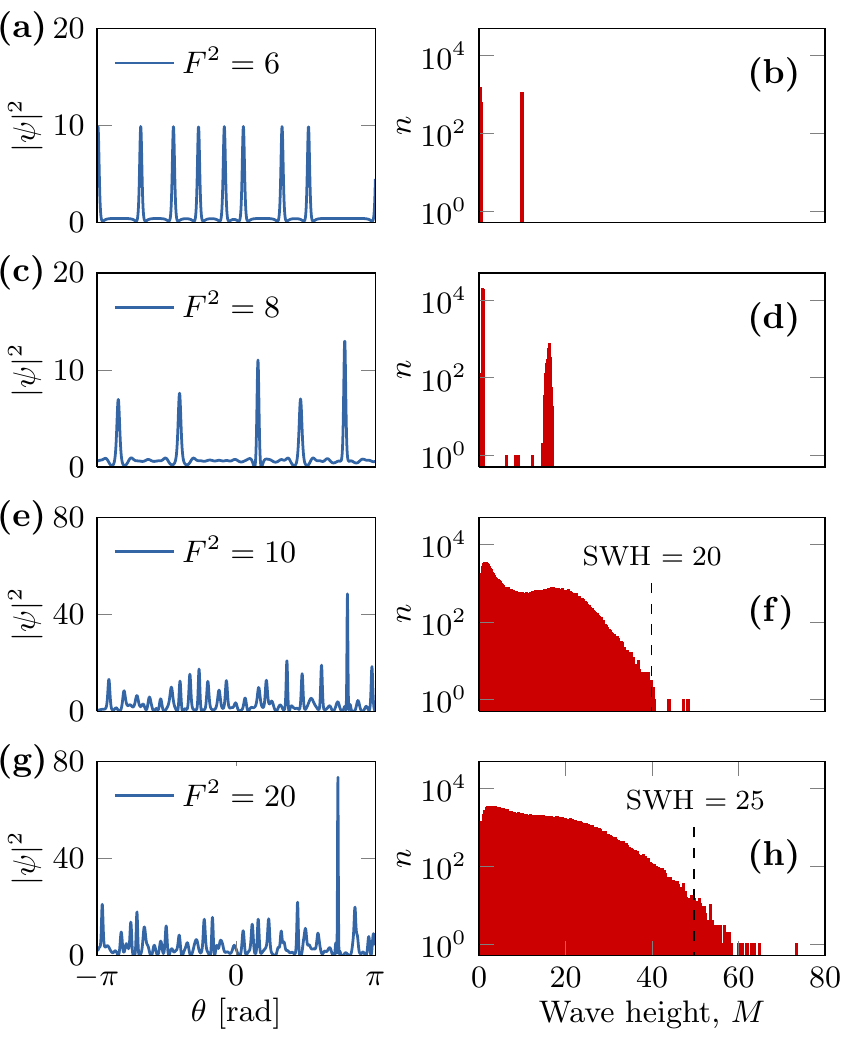}
  \end{center}
  \caption[fig:TempHisto] {(Left column) Spatial distribution of the optical
    intensity in the cavity when the highest wave occurs for different pump
    powers: $F^2=\{6,8,10,20\}$. It should be noted that this spatial
    representation in the moving frame inside the cavity will correspond to a
    temporal signal if the signal is coupled out of the resonator using an output
    coupler. (Right column) The number of events recorded for each wave
    height bin is represented with a logarithmic scale for the vertical axis.
    The local maxima are
    extracted from the maps of Fig.~\ref{fig:pcolors} for each pump power. In
    the cases $F^2=10$ and $F^2=20$, the significant wave height (SWH) is calculated,
    and the black dashed lines correspond to twice this value.}
  \label{fig:TempHisto}
\end{figure}
\begin{figure*}
  \begin{center}
    \includegraphics{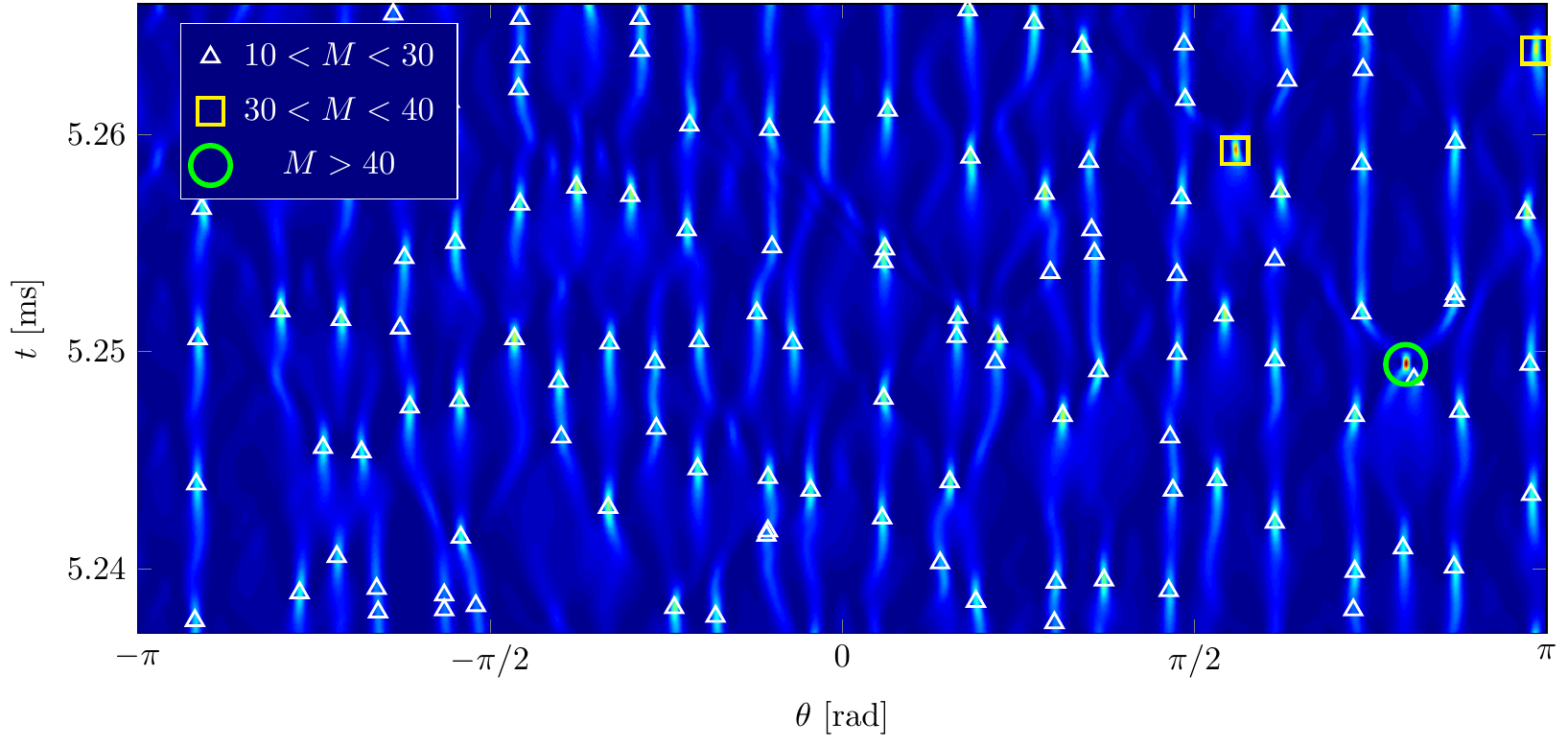}
  \end{center}
  \caption[fig:2Ddynamics]{(Color online) Evolution of the spatial distribution of the optical
  intensity inside the cavity with time. The frequency detuning is $\alpha=4$ and
  excitation is set to $F^2=10$. Triangles are indicating the spatiotemporal location 
  of local maxima whose amplitude $M$ is between $10$ and $30$, corresponding to soliton breathers'
  maxima. The yellow square marks are higher intensity wave laying in the exponential
  tail of the statistical distribution; they correspond to moderate interaction
  between solitons. The green circle marks the highest wave of this
  simulation snapshot, and its extreme amplitude makes it lies outside of the exponential
  tail of the statistical distribution. 
  This unexpected and very strong peak is a sample of what has been identified as a rogue wave in a laser-pumped WGM.}
  \label{fig:2Ddynamics}
\end{figure*}

In optical fibers, extreme events generally arise in the regime of anomalous
dispersion, where multiple bright solitons regime can be observed.
In WGM resonators, bright solitons can also be generated for $\beta<0$ and for
detunings $\alpha$ greater than a critical value which is approximately equal to $2$~\cite{Coillet2013,Coen2013a}.
Here, we report the formation of rogue waves in WGM resonators pumped with a continuous-wave laser.
It is found that pumping the system $10$~times above the threshold for Kerr comb
generation is sufficient to trigger complex chaotic-like (or turbulent) motion for which occurrence of rogue waves has been observed.
Experimentally, this threshold power can be as low as a few
milliwatts, and therefore, the pump power needed to obtain chaotic behavior can
easily be reached~\cite{Chembo2010,Chembo2010a}. 

The LLE simulations are performed using the split-step Fourier method starting
from random noise in the cavity. The WGM resonator's characteristics used for
the simulations are $\beta ={-0.0125}$ and $Q =3 \times 10^9$, corresponding to
a loaded linewidth of $\Delta \omega/2 \pi = 65$~kHz at the standard telecom wavelength
$\lambda = {1550}$~nm. The evolution of the field is simulated over ${10}$~{ms},
a value much higher than the photon lifetime of the cavity $\tau_\text{ph} =
{7.75}$~$\mu$s. 
The transient dynamics is not concerned in our investigation, since we are expecting to obtain rogue waves as rare events of an asymptotic solution of the dynamics. Therefore the first millisecond of the simulation is removed from the time series used for the statistical analysis of the wave height.  The resulting data consists in the
evolution of the field intensity with time for each angle in the azimuthal
direction of the resonator; a color-coded version of a portion of this map is
represented in Fig.~\ref{fig:pcolors}. Each map corresponds to a different value
of the pump power $F^2=\{6,8,10,20\}$, while the detuning is kept constant at
$\alpha=4$.

The next step of our analysis consists in finding the local maxima of the optical field inside the cavity. 
It is worth noting that the minimum pump power leading to Kerr comb generation can theoretically be calculated as 
$F_{\rm th}^2 =1$ for $\alpha =1$ (modulational instability leading to Turing patterns~\cite{Coillet2013}).
Hence, the values of the pump parameters that are considered in the present article ($F^2 <20$) correspond to realistic pump power values ($P<1$~W) that can be easily reached. 
In the first column of Fig.~\ref{fig:TempHisto}, we plot the spatial distribution of the optical field along
the azimuthal direction of the cavity when the highest waves are recorded. On
the second column, the statistical distribution of the wave heights is presented
with a logarithmic scale in the ordinate.
When the pump is fixed to $F^2=6$, solitons emerge from the initial noise in the cavity.
This regime has already been investigated both theoretically~\cite{Coillet2013,Coen2013a}
and experimentally~\cite{Kipp_Solitons}.
In terms of wave height distribution, this regime is characterized by two distinct levels, the soliton
maximum and the small pedestal around them.
When the pump power is increased,
these solitons become unstable and soliton breathers are formed ($F^2=8$)~\cite{Leo2013}.
In this case, the probability distribution of the wave heights is also
given by two peaks histogram corresponding to the maxima of the breathers and
the ripples between them.  When the driving is further increased ($F^2=10$), 
the
soliton breathers 
start to interact one with each other seldom resulting in the occurrence of intense and sharp pulses.
At this point, the statistical
distribution of the peaks height becomes continuous, but still displays the
previously mentioned features corresponding to the breathers' maxima and
ripples. 
The distribution tail for intense events exhibits a specific feature: an exponential decay characterizes the tail up to some threshold  height (dashed line), above which one clearly sees very rare events that qualify for the appellation of ``rogue waves''.

The same behavior is also observed for higher pump power, at $F^2=20$ for
instance, where several extreme events have been observed. The optical intensity
in the cavity when the highest wave occurs is plotted in Fig.~\ref{fig:TempHisto}. 
This wave clearly stands above the others, and results from 
the interaction between previous soliton breathers. The statistical
distribution of this simulation shows that a few events of high intensity indeed
have a higher probability of occurrence than a standard Rayleigh or Gaussian
distribution.

\begin{figure}
  \begin{center}
    \includegraphics{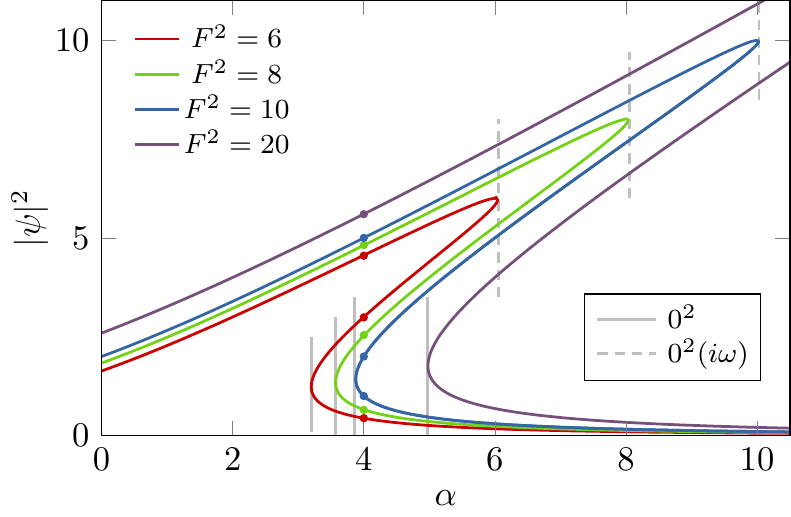}
  \end{center}
  \caption[fig:resonances]
  {(Color online) Nonlinear resonance profiles for different pump powers. Entry and exit from the hysteresis area occur  
                  through $0^2$ and $0^2(i \omega)$ bifurcations respectively. Rogue waves observed in our case are observed for $\alpha =4$ (arbitrary value).}
  \label{fig:resonances}
\end{figure}

A convenient way to characterize rogue waves is to calculate the significant wave height
(SWH), defined as the mean amplitude of the highest third of the waves. An
event will qualify as a rogue wave provided its height is at least two times higher that the
SWH. In the $F^2=10$ and $F^2=20$, the SWH is equal to ${20}$ and~${25}$ respectively,
and a few waves are of greater magnitude in both cases. Rogue waves have
to fulfill this empirical criterion and the field map of Fig.~\ref{fig:pcolors} confirms
that extreme events in the LLE are indeed precisely localized structures in the
spatio-temporal representation. The interpretation of rogue waves as breathers collisions
justifies the short durations that are observed in the simulations. 

From a dynamical point of view, this phenomenology can be explained from the topology of the nonlinear resonance profile of the pumped mode. The angle-independent fixed points $\psi_0$ (or ``flat states'') of the system obey the algebraic equation 
$F^2 = [1+(|\psi_0|^2 -\alpha)^2] \, |\psi_0|^2$,  which is cubic in $|\psi_0|^2$. 
The corresponding resonance profiles are displayed in Fig.~\ref{fig:resonances}.
Rogue waves are observed when the figurative point in the $\alpha$--$F^2$ is located above the $0^2 (i \omega)$ bifurcation line. When the pump power is kept constant, rogue waves are preferably observed before the hysteresis area in the plane   
$\alpha$--$|\psi|^2$. 

In conclusion, we have presented in this work theoretical evidence of 
optical rogue waves in WGM resonators pumped by a continuous-wave laser field.
The spatiotemporal evolution of the intra-cavity field has been modelled
using the Lugiato-Lefever equation. We have focused on
the conditions leading to the generation of extreme events, and the detuning was
chosen in a way such that solitons can emerge from the noisy initial condition. 
As the continuous pump power is increased, extreme events are numerically observed. These events are characterized by shorter duration, amplitudes larger than twice the SWH, and higher probabilities than that of a standard Rayleigh or Gaussian statistics
These
rogue waves therefore appear to emerge from the interaction between cavity soliton breathers, and since this physical system can be controlled with great accuracy, we expect this study to seed a comprehensive understanding of extreme events from a generic standpoint. 

J. M. D. and Y. K. C. acknowledge financial support from the European Research Council 
through the projects ERC AdG MULTIWAVE and ERC StG NextPhase, respectively.

\begin{thebibliography}{19}
\expandafter\ifx\csname natexlab\endcsname\relax\def\natexlab#1{#1}\fi
\expandafter\ifx\csname bibnamefont\endcsname\relax
  \def\bibnamefont#1{#1}\fi
\expandafter\ifx\csname bibfnamefont\endcsname\relax
  \def\bibfnamefont#1{#1}\fi
\expandafter\ifx\csname citenamefont\endcsname\relax
  \def\citenamefont#1{#1}\fi
\expandafter\ifx\csname url\endcsname\relax
  \def\url#1{\texttt{#1}}\fi
\expandafter\ifx\csname urlprefix\endcsname\relax\def\urlprefix{URL }\fi
\providecommand{\bibinfo}[2]{#2}
\providecommand{\eprint}[2][]{\url{#2}}


\bibitem{Review_rogue_waves} N. Akhmediev and E. Pelinovsky (Eds.),        
                             ``Rogue waves - Towards a unifying concept'', 
                              special issue of the Eur. Phys. J. Spe. Top. (2010).

\bibitem[{\citenamefont{Solli et~al.}(2007)\citenamefont{Solli, Ropers,
  Koonath, and Jalali}}]{Solli2007}
\bibinfo{author}{\bibfnamefont{D.~R.} \bibnamefont{Solli}},
  \bibinfo{author}{\bibfnamefont{C.}~\bibnamefont{Ropers}},
  \bibinfo{author}{\bibfnamefont{P.}~\bibnamefont{Koonath}}, \bibnamefont{and}
  \bibinfo{author}{\bibfnamefont{B.}~\bibnamefont{Jalali}},
  \bibinfo{journal}{Nature} \textbf{\bibinfo{volume}{450}},
  \bibinfo{pages}{1054} (\bibinfo{year}{2007}).


\bibitem{Nature-peregrine} B. Kibler, J. Fatome, C. Finot, G. Millot, F. Dias, G. Genty, N. Akhmediev, and J. M. Dudley,
                           Nature Physics \textbf{6}, 790 (2010).
\bibitem{Kovalsky}  M. G. Kovalsky, A. A. Hnilo, and J. R. Tredicce,  
                    Opt. Lett. \textbf{36}, 4449 (2011). 
\bibitem{Bonatto} C. Bonatto, M. Feyereisen, S. Barland, M. Giudici, C. Masoller, J. R. Rios Leite, and J. R. Tredicce,  Phys. Rev. Lett. \textbf{107}, 053901 (2011), 
\bibitem{Pisarchik} A. N. Pisarchik, R. Jaimes-Reátegui, R. Sevilla-Escoboza, G. Huerta-Cuellar, and M. Taki, Phys. Rev. Lett. \textbf{107}, 274101 (2011), 
\bibitem{Arecchi} F. T. Arecchi, U. Bortolozzo, A. Montina, and S. Residori, Phys. Rev. Lett. \textbf{106}, 153901 (2011).

\bibitem[{\citenamefont{Soto-Crespo et~al.}(2011)\citenamefont{Soto-Crespo,
  Grelu, and Akhmediev}}]{SotoCrespo2011}
\bibinfo{author}{\bibfnamefont{J.~M.} \bibnamefont{Soto-Crespo}},
  \bibinfo{author}{\bibfnamefont{P.}~\bibnamefont{Grelu}}, \bibnamefont{and}
  \bibinfo{author}{\bibfnamefont{N.}~\bibnamefont{Akhmediev}},
  \bibinfo{journal}{Phys. Rev. E} \textbf{\bibinfo{volume}{84}},
  \bibinfo{pages}{016604} (\bibinfo{year}{2011}).

\bibitem[{\citenamefont{Zaviyalov et~al.}(2012)\citenamefont{Zaviyalov, Egorov,
  Iliew, and Lederer}}]{Zaviyalov2012}
\bibinfo{author}{\bibfnamefont{A.}~\bibnamefont{Zaviyalov}},
  \bibinfo{author}{\bibfnamefont{O.}~\bibnamefont{Egorov}},
  \bibinfo{author}{\bibfnamefont{R.}~\bibnamefont{Iliew}}, \bibnamefont{and}
  \bibinfo{author}{\bibfnamefont{F.}~\bibnamefont{Lederer}},
  \bibinfo{journal}{Phys. Rev. A} \textbf{\bibinfo{volume}{85}},
  \bibinfo{pages}{013828} (\bibinfo{year}{2012}).

\bibitem[{\citenamefont{Lecaplain et~al.}(2012)\citenamefont{Lecaplain, Grelu,
  Soto-Crespo, and Akhmediev}}]{Lecaplain2012}
\bibinfo{author}{\bibfnamefont{C.}~\bibnamefont{Lecaplain}},
  \bibinfo{author}{\bibfnamefont{P.}~\bibnamefont{Grelu}},
  \bibinfo{author}{\bibfnamefont{J.~M.} \bibnamefont{Soto-Crespo}},
  \bibnamefont{and}
  \bibinfo{author}{\bibfnamefont{N.}~\bibnamefont{Akhmediev}},
  \bibinfo{journal}{Phys. Rev. Lett.} \textbf{\bibinfo{volume}{108}},
  \bibinfo{pages}{233901} (\bibinfo{year}{2012}).

\bibitem{Review_John} N. Akhmediev, J. M. Dudley, D. R. Solli, and S. K. Turitsyn,
                      J. Opt. \textbf{15}, 060201 (2013)


\bibitem{Matsko} A. B. Matsko, A. A. Savchenkov, W. Liang, V. S. Ilchenko, D. Seidel, and L. Maleki,
                      Opt. Lett. \textbf{36}, 2845 (2011).

\bibitem[{\citenamefont{Chembo and Menyuk}(2013)}]{Chembo2013}
\bibinfo{author}{\bibfnamefont{Y.~K.} \bibnamefont{Chembo}} \bibnamefont{and}
  \bibinfo{author}{\bibfnamefont{C.~R.} \bibnamefont{Menyuk}},
  \bibinfo{journal}{Phys. Rev. A} \textbf{\bibinfo{volume}{87}},
  \bibinfo{pages}{053852} (\bibinfo{year}{2013}).

\bibitem[{\citenamefont{Coen et~al.}(2013)\citenamefont{Coen, Randle,
  Sylvestre, and Erkintalo}}]{Coen2013}
\bibinfo{author}{\bibfnamefont{S.}~\bibnamefont{Coen}},
  \bibinfo{author}{\bibfnamefont{H.~G.} \bibnamefont{Randle}},
  \bibinfo{author}{\bibfnamefont{T.}~\bibnamefont{Sylvestre}},
  \bibnamefont{and}
  \bibinfo{author}{\bibfnamefont{M.}~\bibnamefont{Erkintalo}},
  \bibinfo{journal}{Opt. Lett.} \textbf{\bibinfo{volume}{38}},
  \bibinfo{pages}{37} (\bibinfo{year}{2013}).

\bibitem[{\citenamefont{Lugiato and Lefever}(1987)}]{Lugiato1987}
\bibinfo{author}{\bibfnamefont{L.~A.} \bibnamefont{Lugiato}} \bibnamefont{and}
  \bibinfo{author}{\bibfnamefont{R.}~\bibnamefont{Lefever}},
  \bibinfo{journal}{Phys. Rev. Lett.} \textbf{\bibinfo{volume}{58}},
  \bibinfo{pages}{2209} (\bibinfo{year}{1987}).

\bibitem[{\citenamefont{Coillet et~al.}(2013)\citenamefont{Coillet, Balakireva,
  Henriet, Saleh, Larger, Dudley, Menyuk, and Chembo}}]{Coillet2013}
\bibinfo{author}{\bibfnamefont{A.}~\bibnamefont{Coillet}},
  \bibinfo{author}{\bibfnamefont{I.}~\bibnamefont{Balakireva}},
  \bibinfo{author}{\bibfnamefont{R.}~\bibnamefont{Henriet}},
  \bibinfo{author}{\bibfnamefont{K.}~\bibnamefont{Saleh}},
  \bibinfo{author}{\bibfnamefont{L.}~\bibnamefont{Larger}},
  \bibinfo{author}{\bibfnamefont{J.}~\bibnamefont{Dudley}},
  \bibinfo{author}{\bibfnamefont{C.}~\bibnamefont{Menyuk}}, \bibnamefont{and}
  \bibinfo{author}{\bibfnamefont{Y.~K.}~\bibnamefont{Chembo}},
  \bibinfo{journal}{IEEE Photonics Journal} \textbf{\bibinfo{volume}{5}},
  \bibinfo{pages}{6100409} (\bibinfo{year}{2013}).

\bibitem[{\citenamefont{Coen and Erkintalo}(2013)}]{Coen2013a}
\bibinfo{author}{\bibfnamefont{S.}~\bibnamefont{Coen}} \bibnamefont{and}
  \bibinfo{author}{\bibfnamefont{M.}~\bibnamefont{Erkintalo}},
  \bibinfo{journal}{Opt. Lett.} \textbf{\bibinfo{volume}{38}},
  \bibinfo{pages}{1790} (\bibinfo{year}{2013}).

\bibitem[{\citenamefont{Chembo et~al.}(2010)\citenamefont{Chembo, Strekalov,
  and Yu}}]{Chembo2010}
\bibinfo{author}{\bibfnamefont{Y.~K.} \bibnamefont{Chembo}},
  \bibinfo{author}{\bibfnamefont{D.~V.} \bibnamefont{Strekalov}},
  \bibnamefont{and} \bibinfo{author}{\bibfnamefont{N.}~\bibnamefont{Yu}},
  \bibinfo{journal}{Phys. Rev. Lett.} \textbf{\bibinfo{volume}{104}},
  \bibinfo{pages}{103902} (\bibinfo{year}{2010.

\bibitem[{\citenamefont{Chembo and Yu}(2010)}]{Chembo2010a}
\bibinfo{author}{\bibfnamefont{Y.~K.} \bibnamefont{Chembo}} \bibnamefont{and}
  \bibinfo{author}{\bibfnamefont{N.}~\bibnamefont{Yu}}, \bibinfo{journal}{Phys.
  Rev. A} \textbf{\bibinfo{volume}{82}}, \bibinfo{pages}{033801}
  (\bibinfo{year}{2010}).

\bibitem{Kipp_Solitons} T.~Herr, V~Brasch, J.~D.~Jost, C.~Y.~Wang, N.~M.~Kondratiev, M.~L.~Gorodetsky, T.~J.~Kippenberg,
           \emph{arXiv:1211.0733v3}, June~2013.              

\bibitem[{\citenamefont{Leo et~al.}(2013)\citenamefont{Leo, Gelens, Emplit,
  Haelterman, and Coen}}]{Leo2013}
\bibinfo{author}{\bibfnamefont{F.}~\bibnamefont{Leo}},
  \bibinfo{author}{\bibfnamefont{L.}~\bibnamefont{Gelens}},
  \bibinfo{author}{\bibfnamefont{P.}~\bibnamefont{Emplit}},
  \bibinfo{author}{\bibfnamefont{M.}~\bibnamefont{Haelterman}},
  \bibnamefont{and} \bibinfo{author}{\bibfnamefont{S.}~\bibnamefont{Coen}},
  \bibinfo{journal}{Opt. Express} \textbf{\bibinfo{volume}{21}},
  \bibinfo{pages}{9180} (\bibinfo{year}{2013}).

%
%
%
%
%


}

\end{thebibliography}


\end{document}